\documentclass[11pt,a4paper,reqno]{amsart}
\usepackage{amsmath,amsthm,amsfonts,amssymb,bm} 
\usepackage{amsaddr}
\usepackage{comment} 
\usepackage{bbm} 
\usepackage[headings]{fullpage}
\usepackage{graphicx} 
\usepackage{hyperref} 
\usepackage[symbol]{footmisc}

\theoremstyle{plain}
\newtheorem{theorem}{Theorem}
\newtheorem{proposition}[theorem]{Proposition}
\newtheorem{corollary}[theorem]{Corollary}
\newtheorem{lemma}[theorem]{Lemma}

\theoremstyle{definition}

\theoremstyle{remark}

\def\d{{\rm d}}

\def\E{\mathbb E}

\def\su{\subseteq}

\def\mc{\mathcal}
\def\ms{\mathsf}

\def\Q{\mathbb Q}
\def\P{\mathbb P}

\def\R{\mathbb R}

\def\a{\alpha}
\def\s{\sigma}
\def\d{\mathrm d}

\def\lan{\langle}
\def\ran{\rangle}
\def\bel{\begin{lemma}}
\def\enl{\end{lemma}}
\def\bep{\begin{proof}}
\def\enp{\end{proof}}
\def\mc{\mathcal}
\def\LL{\mc L}
\def\MM{\mc M}
\def\de{\delta}
\def\De{\Delta}
\def\bepr{\begin{proposition}}
\def\enpr{\end{proposition}}
\def\bet{\begin{theorem}}
\def\ent{\end{theorem}}
	\def\bec{\begin{corollary}}
	\def\enc{\end{corollary}}
\def\e{\varepsilon}
\providecommand{\lf}[1]{\lan f,#1\ran}

\def\been{\begin{enumerate}}
\def\enen{\end{enumerate}}

	\def\th{\theta}

\def\xis{\xi \in \CC}

\def\DT{D([0, T], \MM(S))}
\def\DTT{\MM(D_T)}
\def\ff{\infty}
\def\CC{\mc C}

\def\barmutx{\bar{\mu}_{t, \xi}}

\def\bs{\boldsymbol}
\def\le{\leqslant}
\def\ge{\geqslant}
\def\f{\frac}
\def\t{\tau}

\begin{document}
\title[Artificial neural networks with sparse connectivity]{Asymptotic properties of one-layer artificial neural networks with sparse connectivity}
	
\author[Hirsch, Neumann, Schmidt]{Christian Hirsch$^{1,2}$, Matthias Neumann$^3$, Volker Schmidt$^3$}

\address{$^1$Bernoulli Institute for Mathematics, Computer Science and Artificial Intelligence, University~of~Groningen, Nijenborgh 9, NL-9747AG Groningen, Netherlands}
\address{$^2$CogniGron (Groningen Cognitive Systems and Materials Center), University of Groningen, Nijenborgh 4, NL-9747AG Groningen, Netherlands}
\address{$^3$Institute of Stochastics, Ulm University, Helmholtzstra\ss e 18, 89069 Ulm, Germany}

\email{hirsch@au.dk, matthias.neumann@uni-ulm.de, volker.schmidt@uni-ulm.de}


\thanks{\textit{Acknowledgements.} CH acknowledges financial support of the CogniGron research center and the Ubbo Emmius Funds (University of Groningen). The work of MN was partially funded by the POLiS Cluster of Excellence (EXC~2154/1). }

\begin{abstract}
A law of large numbers for the empirical distribution of parameters of a one-layer artificial neural networks with sparse connectivity is derived for a simultaneously increasing number of both, neurons and training iterations of the stochastic gradient descent.
\end{abstract}

\maketitle

\section{Introduction}
\label{sec:intro}
Machine learning and artificial neural networks in particular, are shaping the future of science by providing powerful tools for a data-driven gain of knowledge. The simplest architecture of an artificial neural network (ANN) is given by a \emph{single-layer perceptron} (SLP), i.e., a feed-forward network with one layer~\cite{hastie} and fully connected neurons. For \emph{deep learning}, where ANNs with more than one layer are considered, fully connected layers are still indispensable \cite{goodfellow}. However, connections between neurons in biological neural networks are typically sparse~\cite{pessoa}. This inspired the development of ANNs with sparse connectivity between neurons, which exhibit -- in terms of accuracy -- the same quality as their fully connected counterparts~\cite{mocanu}. In the present paper, we provide a theoretical analysis of SLPs with sparse connectivity, which are trained via  stochastic gradient descent (SGD)~\cite{goodfellow}. By extending the methods considered in~\cite{sirignano.2018}, we derive a law of large numbers (LLN) for the empirical distribution of parameters for the asymptotic regime, where both, the number of neurons and training iterations of the SGD are simultaneously increasing. We consider a model with random sparsity~\cite{kaviani}, which is -- in contrast to the adaptive approach considered in~\cite{mocanu} -- pre-defined before training~\cite{dey}. Connections between input data and the different neurons in the hidden layer are removed independently. The considered model particularly covers the Erd\H{o}s-R\'enyi graph, which serves as the initial state for the adaptive connectivity model in~\cite{mocanu}. The formal definition of the ANN model considered in the present paper as well as the main results are given in Section~\ref{sec:main_results}. Subsequently, in Section~\ref{sec:simulation}, the main results are illustrated by means of a simulation study. The rest of the paper is dedicated to the proofs, where we follow the basic idea of~\cite{sirignano.2018} and consider the development of the empirical distribution of the ANN-parameters as an element in an appropriately chosen Skorokhod space. Then, weak convergence of these objects in the asymptotic regime mentioned above is obtained by building on a blueprint that has already been successfully implemented in a variety of contexts such as those considered in~\cite{dc,simon,sirignano.2020}. More precisely, in our case, we show tightness of the sequence under consideration in the Skorokhod space (Section~\ref{tight_sec}), uniqueness of the limit (Section~\ref{uniq_sec}) and identify the limit (Section~\ref{identify_sec}).


\section{Definitions and main results}
\label{sec:main_results}
Our approach is based on results presented in~\cite{sirignano.2018} which investigates asymptotic properties of an SLP consisting of an input layer with $d \ge 1$ nodes and one hidden layer of $N \ge 1$ nodes. More precisely, let $x \in \R^d$ be the input vector, and  $c^1, \dots, c^N \in \R$, $w^1, \dots, w^N \in\R^d$ be the weights of the SLP for the output and hidden layer, respectively. Denoting by 
$$\bs \th  = (c^1, \dots, c^N, w^1, \dots, w^N) \in \R^{(1 + d)N}$$ 
the vector containing all weights, the SLP $g(x, \bs \th)$ with parameter $\th$ is defined by
\begin{equation}\label{eq:ann}
		g(x, \bs \th) = \frac1N \sum_{i \le N} c^i \s ( x^\top w^i  ),
\end{equation}
where we assume that the activation function $\s:\, \R \rightarrow \R$ is a twice differentiable bounded function with bounded derivatives.

%
%
Formalizing the setup of \cite{mocanu}, we modify the above SLP such that for $1 \le i \le N,$ the $i$th node in the hidden layer is influenced only by a certain subset $\xi^i \su \{1, \dots, d\}$ of the coordinates of the input vector.  Thus, for each $1\leq i \leq N$, we put those coordinates of $w^i $ equal to 0 that do not belong to $\xi^i$.
Depending on the application context, it may make sense to select $\xi^i$ only from a subset of \emph{admissible prunings} $\CC \subset \{A:\, A\su \{1, \dots, d\} \}$, which is fixed henceforth.  An essential example corresponds to the setting, where the $\{\xi^i\}_{i \ge 1}$ are realizations of independent and identically distributed (iid) configurations $\{\Xi^i\}_{i \ge 1}$. For instance, in the simulation study described in Section~\ref{sec:simulation}, we consider \emph{Erd\H{o}s-R\'enyi pruning} with parameter $0< p \le 1$, where $\P(\Xi^1 = \xi)=  p^{\#\xi} (1 - p)^{d - \#\xi}$. 

%
%
Now, let $\{(X_k,Y_k)\}_{k \ge 1}$ be a random sequence of iid training data, where for each $k \geq 1$, the random vector $(X_k, Y_k)$ is a copy of a random vector $(X,Y):\Omega \rightarrow \R^{d+1}.$ Then, we train the SLP through SGD with respect to the squared-error loss function $(x, y) \mapsto (y - g(x, \th))^2$ and learning rate $\a_N = \a / N$ for some $\a > 0$. More precisely,  we initialize the network with random weights $\bs \th_0$ and then iteratively update them via
	\begin{equation}
				\begin{aligned}
						\label{eq:sgd}
									c^i_{k+1} &= c_k^i + \frac1Ng(X_k, Y_k, \bs \th_k) \, \s (X^\top_k w_k^i   ), \\
												w^i_{k+1} &= w_k^i + \frac1Ng(X_k, Y_k, \bs \th_k)\, c_k^i \, \s^\prime (X^\top_k  w_k^i ) X_k(\xi^i),
				\end{aligned}  
					\end{equation}
					where 
					$g(X_k, Y_k, \bs \th_k) = \a(Y_k - g(X_k,\bs \th_k))$ 
					and $X_k(\xi^i)$ denotes the modification of $X_k$ with entries of $X_k$ outside $\xi^i$ set to 0. 

					%
					%
					The main result of the present paper describes the  evolution of the parameter $\bs \th$ if the number of SGD iterations is of order $N$. Our key innovation to the analysis in comparison to \cite{sirignano.2018} is that due to the recursion given in \eqref{eq:sgd}, where weights corresponding to different $\xi^i$ evolve differently. Hence, when understanding the evolution over time, these groups of weights need to be separated. As a result, we obtain a law of large numbers, which is quenched on the $\xi$-configuration.

					The main idea to arrive at the quenched LLN is to choose a tailormade state space that allows for a smooth extension of the argument used in \cite{sirignano.2018}.
					More precisely, let $S_\xi = \R^{1 + d}$ be a separate copy of $\R^{1 + d}$ for each $\xis$, and let 
					$S= \bigsqcup_{\xis}  S_\xi,$
					be the disjoint union of these copies. In this set-up the $i$th weight vector $\th^i$ is considered to be embedded inside $S_{\xi^i} \su S$.
					Moreover, a function $f:\, S \to \R$ corresponds to a collection of functions $f = \{f_\xi\}_{\xis}$ defined on each $S_\xi$. For each $\xis$, a probability measure $\mu$ on $S$ defines  a probability measure on $S_\xi$ via 
					$\mu_\xi(\cdot) =  \mu(\cdot) / \mu(S_\xi).$

					%
					%
					In this interpretation, we let
					\begin{equation*}
							\nu_k^N = \frac1N \sum_{i \le N} \de_{\th_k^i}
					\end{equation*}
					denote the empirical measure of the weights after $k \ge 1$ iterations. In particular, $\nu_k^N$ is a random element in the space $\MM(S)$ of probability measures on $S$.
					We interpret 
					$$g(X_k, \nu_k^N) = \langle g(X_k, \cdot), \nu_k^N\rangle = \int_Sg(X_k, \th) \nu_k^N(\d \th)$$
					as the integration of the function
					$
							g(X_k, \cdot)\colon\R^{1 + d} \to \R$, $(c, w) \mapsto c \s(X_k^\top w)$					with respect to $\nu_k^N$. A similar remark holds for
				$g(X_k, Y_k, \nu_k^N)$. 
					%
					%
					Then, we show that as $N \to \ff$, the time-rescaled measure 
$$\mu_t^N = \nu_{\lfloor Nt \rfloor}^N$$
					converges to the solution of an evolution equation described in \eqref{evol_eq} below. We think of $\mu^N$ as a random element in the Skorokhod space $D([0, T], \MM(S))$. For the rest of the paper, we fix $p_\xi > 0$, $\xis$ with $\sum_{\xis} p_\xi = 1$ and assume that
					\begin{itemize}
					\item[{\bf (E)}] 
					 $\lim_{N \to \infty} \, \frac{1}{N} \, \#\{ i\le N:\, \xi^i = \xi\} = p_\xi$ (ergodicity condition),
					    \item[{\bf (M)}]  the random sequences of the initial parameters $\{c_0^i\}_{i \le N}$ and $\{w_0^i\}_{i \le N}$ are both iid, independent of each other, and satisfy  $\E[\exp(q |c^i_0|) + |w^i_0|^4] < \ff$ for some $ q > 0$. Moreover,   $\E [|X|^6 + Y^6] < \infty$ (moment condition).
					    
					    \end{itemize}

					 %
					 %
					 \bet[Quenched LLN]
					 	\label{thm:lln}
							Under the conditions {\bf(E)} and  {\bf(M)}, the limit trajectory $\bar\mu_\cdot = \lim_{N \rightarrow \infty} \mu^N_\cdot $ exists and decomposes as 
								\begin{align}
											\label{dec_eq}
													\bar\mu_t = \sum_{\xis} p_\xi \barmutx.
														\end{align}
														Moreover, for each $f \in C_b^2(S)$, the trajectory $\{\bar\mu_t\}_{t \le T}$ satisfies
															\begin{align}
																	    \label{evol_eq}
																	    		\frac\d{\d t}\lan f, \bar\mu_t \ran =  \lan A(\cdot; \bar\mu_t)\nabla f , \bar\mu_t\ran
																				\end{align}
																					with $A(\th; \bar \mu_t) = \big(A_{\ms c}(\th;\bar\mu_t), A_{\ms w}(\th;\bar\mu_t)\big)$, where
																							\begin{align*}
																								A_{\ms c}(\th;\bar\mu_t) &= \E\big[g(X, Y, \bar\mu_t)  \s(X^\top w ) \big],\, \, \\
																								A_{\ms w}(\th; \bar\mu_t) &= \E\big[g(X, Y, \bar\mu_t)c \s'(X^\top w )X(\xi) \big] \quad\text{ if $\th \in S_\xi \subseteq S$}.
																							\end{align*}
\ent


\section{Simulation study}\label{sec:simulation}
We illustrate the law of large numbers stated in Theorem~\ref{thm:lln} by means of a simulation study. For this purpose, we approximate the function $f:[0,1]^2 \rightarrow \R$ defined by $f(s,t) = \sin(st) \sqrt{\log(1+t)} + \cos(t^2), (s,t) \in [0,1^2]$ by the SLP $g(x,\bs \th)$ after Erd\H{o}s-R\'enyi pruning with parameter $p=1/2$ as defined in Section~\ref{sec:main_results}. The activation function is chosen to be $\sigma:\R \rightarrow \R$ with  $\sigma(s) = (1 - \exp(-t)) / (2 + 2 \exp(-t))$. Training is performed via SGD as given in \eqref{eq:sgd} with learning rate $\alpha = 100$ and the number of iterations is chosen to be $KN$, where we put $K=1,000.$ As training data, we consider collections of random vectors $(X_1, f(X_1)), \ldots, (X_{KN},f(X_{KN}))$, where $X_1, \ldots, X_{KN}$ are independent and uniformly distributed on the unit square, {i.e.}, $X_i \sim U([0,1]^2)$ for each $i\in\lbrace 1,\ldots,KN \rbrace$. The initial parameter configuration is chosen at random, where each of the sequences $c^1_0, \ldots, c^N_0$ and $w^1_0, \ldots, w^N_0$ is iid with $c^1_0 \sim U(-10,10), w^1_0 \sim U([-10,10]^2)$.
\begin{figure}[h!]
    \centering
    \includegraphics[width = 0.99\textwidth]{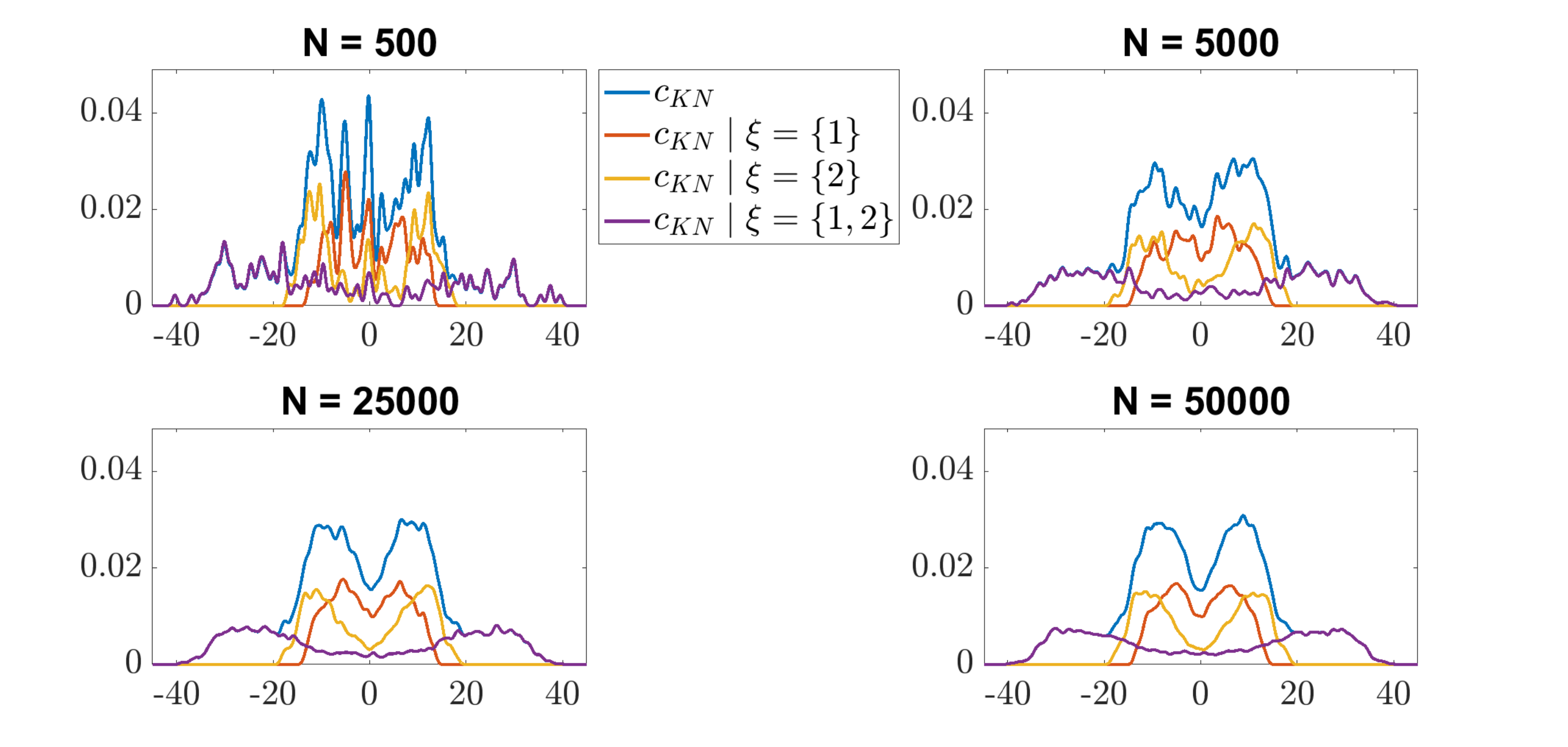}
    \caption{Probability density function of the real-valued parameter $c_{KN}$ after $KN$ iterations of the SGD for $N \in \lbrace 500, 5000, 25000, 50000 \rbrace,$ obtained via kernel density estimation. The distribution is a mixture of the conditional distributions $c_{KN}$ given that $\Xi = \xi$, where $\xi$ is a subset of $\lbrace 1,2 \rbrace,$ which are also shown in each of the four plots. Note that conditional probability density functions are scaled such that their sum gives the (unconditional) probability density functions of $c_{KN}$.}
    \label{fig:distribution_parameter_c}
\end{figure}

For $N \in \lbrace 500, 5000, 25000, 50000 \rbrace$,  
Figure~\ref{fig:distribution_parameter_c} shows the empirical distribution of the sample $c^1_{KN}, \ldots, c^N_{KN}$ in terms of probability density functions which are obtained by kernel density estimation  with a Gaussian kernel and a fixed bandwidth of $0.5$. Moreover, the empirical distributions conditioned on the realization of $\Xi$ (defining the network topology as described in Section~\ref{sec:main_results}) are shown. This illustrates clearly that the distribution of $c_{KN}$ is a mixture of the distributions conditioned on realizations of $\Xi$. Figure~\ref{fig:distribution_parameter_c} shows the convergence of the distribution of $c_{KN}$. Only minor changes in the distribution can be observed between $N = 25000$ and $N=50000$. Additionally, the empirical bivariate distributions of the samples $(c^1_{KN}, (w^1_{KN})_1), \ldots, (c^N_{KN}, (w^N_{KN})_1)$, $(c^1_{KN}, (w^1_{KN})_2), \ldots, (c^N_{KN}, (w^N_{KN})_2)$, and $((w^1_{KN})_1), (w^1_{KN})_2), \ldots, ((w^1_{KN})_1), (w^N_{KN})_2)$ are provided as supplementary material.

\section{Outline of proof}
\label{sec:proofs}
As in \cite{sirignano.2018}, we pursue the well-established three-step procedure for weak convergence towards a limiting process, which has been implemented in \cite{dc,simon}. We now state the three steps in detail and observe that they indeed imply the asserted Theorem \ref{thm:lln}. The proofs of the three results are deferred to Sections \ref{tight_sec}, \ref{uniq_sec} and \ref{identify_sec}.

%
%
\bepr[Tightness]
\label{tight_prop}
Under conditions {\bf (E)} and {\bf (M)} the sequence $\{\LL(\mu^N)\}_{N \ge 1}$ of distributions of the measures $\mu^N$ is tight.
\enpr

%
%
\bepr[Uniqueness]
\label{uniq_prop}
For a given initial value and given $\{p_\xi\}_{\xis}$ with $\sum_{\xis} p_\xi = 1$, Equation \eqref{evol_eq} has at most one solution $\bar\mu_t$ with $\bar\mu_t(S_\xi) = p_\xi$.
\enpr

%
%
\bepr[Limit identification]
\label{identify_prop}
Under condition {\bf (E)}, any weak accumulation point of $\{\LL(\mu^N)\}_{N \ge 1}$ satisfies Equation \eqref{evol_eq}.
\enpr

%
%
To make the presentation self-contained, we formally conclude the proof of Theorem~\ref{thm:lln}.
\bep[Proof of Theorem~\ref{thm:lln}]
First, under condition {\bf (E)},
$\mu_t^N(S_\xi) = \frac1N\#\{ i\le N:\, \xi^i = \xi\}$
converges to $p_\xi$, thereby yielding the decomposition \eqref{dec_eq}. Next, by Proposition \ref{tight_prop}, any subsequence of $\{\LL(\mu^N_\cdot)\}_{N \ge 1}$ has a weakly convergent subsequence. By Propositions \ref{uniq_prop} and \ref{identify_prop}, any such subsequence converges weakly to the unique solution of \eqref{evol_eq}. Hence, also the entire sequence $\{\LL(\mu^N_\cdot)\}_{N \ge 1}$ converges in distribution to that solution.
\enp

\section{Tightness}
\label{tight_sec}
In this section, we show tightness of the sequence $\{\LL(\mu^N)\}_{N \ge 1}$ in the Skorokhod space $\DT$. To that end, we rely on the established method, which involves compact containment and regularity, see Theorem 4.5 in~\cite{jacod}. In particular, the following assertions are true.

%
%
\bepr[Compact containment]
\label{cc_lem}
Let $\e > 0$.Then, for some compact $K \su S$,
$$\sup_{N \ge 1}\sup_{t\le T}\P(\mu_t^N \not \in K) \le \e.$$
\enpr

%
%
For regularity, we rely on Aldous' celebrated criterion, see Lemma 16.12 in~\cite{kallenberg.2002}. 
\bepr[Aldous' criterion]
\label{reg_lem} 
Let $f \in C_b^2(S_\xi)$. Then,
\begin{equation}\label{eq:aldous_criterion}
    \lim_{\delta \rightarrow 0} \limsup_{N \to \ff}\sup_\t\P\big(\sup_{u \le \de}|\lf{\mu_{\t + u}^N}- \lf{\mu_\t^N} | \ge \e\big) = 0,
\end{equation}
where, $\t$ is taken from the family of all stopping times that are bounded by $T$.
\enpr

Note that in order to verify~\eqref{eq:aldous_criterion} it is sufficient to show that
$$\lim_{\delta \rightarrow 0} \limsup_{N \rightarrow \infty} \sup_{\sigma, \tau} \E \left[   |\lf{\mu_{\sigma}^N}- \lf{\mu_\t^N}| \wedge  1   \right] = 0,$$ where $\tau$ and $\sigma$ are taken from the family of stopping times fulfilling  $\sigma \leq \tau \leq \sigma + \delta \le T.$
The proof of Proposition~\ref{cc_lem} is analogous to that of Lemma 2.2 in \cite{sirignano.2018}. Hence, we focus on Proposition~\ref{reg_lem}, where the arguments that we present differ from those used in \cite{sirignano.2018}.

First, we bound the increments of the parameters during SGD.  To that end, we rewrite \eqref{eq:sgd} succinctly as
\begin{align}											         \label{eq:sgd_succ}
\th_{k + 1}^i - \th_k^i = \f1N B_k^N(\th_k^i),
\end{align}
where $B_k^N(\th) = \big(B_{k, \ms c}(\th), B_{k, \ms w}(\th)\big)$ with
$
	B_{k, \ms c}^N(\th) = g(X_k, Y_k, \nu^N_k) \sigma(w^\top X_k)
	$
and
\begin{align*}
B_{k, \ms w}^N(\th) = g(X_k, Y_k, \nu^N_k)c \sigma'(w^\top X_k)X_k(\xi)\text{ if $\th \in S_\xi$}.
\end{align*}

To prove Proposition  \ref{reg_lem}, we first discuss an auxiliary result. 
Instead of directly bounding the parameters as in Lemma~2.1 of~\cite{sirignano.2018}, we found it more convenient to concentrate on the increments. As a preliminary step, we also rely on a related property for independent random variables, which we state and prove here to make the presentation self-contained.
%
%
\bel[Regularity for independent random variables]
\label{lemma:mod_cont}
Let $\{Z_k\}_{k \ge 1}$ be a family of iid non-negative random variables with finite second moment. Then, as $\delta$ tends to 0,  
$$\limsup_{N \to \infty}\f1N\E\Big[\max_{{ k \le N}}\sum_{{k \le \ell  \le k+ \de N}} Z_\ell\Big] \in O(\de).$$
\enl
\bep
First, we expand the expression under the expectation as
$$\max_{k\le N}\sum_{\ell = k}^ { k+ \de N} Z_\ell \le \max_{m \le 1/ \de}\sum_{\ell = m\de N}^{  m\de N+ 2\de N} Z_\ell= 2\de N\E Z_1  + \sqrt N\max_{m \le 1/ \de}\sum_{\ell = m\de N}^{ m\de N+2\de N} \frac{Z_\ell - \E Z_\ell }{\sqrt N}.$$
Since $\sqrt N \in o(N)$, it suffices to show that the second moment of the above sum is bounded for each $m$. Now, leveraging independence, we get that
\begin{align*}
	\ms{Var}\Big(\sum_{\ell = m\de N}^{m\de N+ 2\de N}\frac{Z_\ell - \E Z_\ell }{\sqrt N}\Big) = 2\de \ms{Var} Z_1  < \ff.&\qedhere
\end{align*}
\enp
\bel[Boundedness of increments]
\label{lemma:bound_c_w}
Assume condition ${\mathbf{(M)}}$. Then, as $\delta$ tends to $0$, 
$$\limsup_{N \to \ff}\frac1N \E\Big[\max_{{ k \le NT}}\sum_{{k \le \ell  \le k+ \de N}} \lan |B_\ell^N(\cdot)|^2, \nu_\ell^N\ran\Big] \in O(\de).$$
\enl
\bep

%
%

 We deal with the $c_k^i$- and $w_k^i$- increments separately. 
First, according to \eqref{eq:sgd_succ} and the boundedness of $\s$, there are constants $C_1, C_1' > 0$ such that for every $k \le \ell \le k'$, 
	\begin{align}
	\label{c_ell_eq_preliminary}
		|c^{i}_{\ell+1}| \le |c^i_\ell| +
	\f{C_1}N	 |Y_\ell| + \f{C_1}N|g(X_\ell, \nu_\ell)| \le |c^i_\ell| +  \f{C_1}N|Y_\ell| + \frac{C_1'}{N^2} \sum_{j \le N} | c_\ell^j |.
	\end{align}
	Hence, writing $\overline Y_N = \tfrac 1N \sum_{\ell \le N}|Y_\ell|$, 
	we argue as in \cite[p.734, line -1]{sirignano.2018} to show that 
	\begin{align}
		\label{c_ell_eq}
	|c_\ell^i| \le C_2\Big(|c_0^i| + \f1N\sum_{j \le N}|c_0^j| + \overline Y_N \Big),
	\end{align}
	for some $C_2 > 0$.
	In particular, we can find a suitable $C_3 > 0$ such that 
	$
		\lan B_{\ell, \ms c}^N(\cdot)^2, \nu_\ell^N\ran \le  C_3\big(Y_\ell^2 + \overline Y_N^2 + \overline C_N^2 \big),
		$
	where $\overline C_N^2 = N^{-1} \sum_{j \le N}(c_0^j)^2$.
	Thus,
 Lemma \ref{lemma:mod_cont} yields the claim for the $c_k^i$-increments. 
  Similarly, for the $w_k^i$-increments, the bound \eqref{c_ell_eq} yields suitable constants $C_4, C_4' > 0$ such that
\begin{align*}
	\big|B_{\ell, \ms w}^N(\th_\ell^i)\big| \le C_4\big( |Y_\ell| + \f1N\sum_{j \le N} | c_\ell^j | \big)|X_\ell||c_\ell^i| \le C_4'
\big(|Y_\ell| + \f1N\sum_{j \le N}|c_0^j| + \overline Y_N \big)|X_\ell||c_\ell^i|.
	\end{align*}
	In particular, applying \eqref{c_ell_eq} and using $abc \le (a^3 + b^3 + c^3)/3$ for $a,b,c >0$, we get that
	$$
	\lan |B_{\ell, \ms w}^N(\cdot)|^2, \nu_\ell^N\ran \le C_5\big(Y_\ell^2 + \overline C_N^2 + \overline Y_N^2  \big)\f{|X_\ell|^2}N\sum_{i \le N}|c_\ell^i|^2\\
	\le C_5'\big(Y_\ell^6 + |X_\ell|^6 +\overline C_N^6 +  \overline Y_N^6  \big)
	$$
for suitable $C_5, C_5' > 0$. Therefore, 
\begin{align*}
	&\sum_{k \le \ell \le k'}	\lan |B_{\ell, \ms w}^N(\cdot)|^2, \nu_\ell^N\ran \le C_5' \sum_{k \le \ell \le k'}(Y_\ell^6  + |X_\ell|^6) + (k' - k)\overline C_N^6 +  (k' - k)\overline Y_N^6  ,
	\end{align*}
	so that an application of Lemma \ref{lemma:mod_cont} concludes the proof.
\enp

%
%
Finally, we prove Proposition  \ref{reg_lem}.

%
%
\bep[Proof of Proposition \ref{reg_lem}]
To ease notation, we omit henceforth the $\lfloor \cdot \rfloor$-symbols and write $Ns$ instead of $\lfloor Ns \rfloor$. In particular, we write $\mu_t^N =  \nu_{Nt}^N$.
 Then,
by Taylor expansion, we find intermediate values $\{\bar\th_k^i\}_{i \ge 1} \su S$ such that	
\begin{align*}
|\lf{\nu_{N(\t + u)}^N} - \lf{\nu_{N\t}^N}| 
	&\le \f1N\sum_{N\t \le \ell \le N(\t + u)}\Big|\lan B_\ell^N(\cdot)\nabla f, \nu_\ell^N\ran\Big| \\
	&\phantom{\le}+\frac 1{2N^2}\sum_{i \le N}\sum_{N\t \le \ell \le N(\t + u)}   \Big|B_\ell^N(\th_\ell^i)\nabla^2f(\bar\th_\ell^i)B_\ell^N(\th_\ell^i)^\top\Big|.
\end{align*}
By assumption, all first- and second-order partial derivatives of $f$ are uniformly bounded, which means that there exist $C_1, C_2 > 0$ such that 
\begin{align*}
|\lf{\nu_{N(\t + u)}^N} - \lf{\nu_{N\t}^N}| 
	&\le \f {C_1}N\sum_{N\t \le \ell \le N(\t + u)}\big(\lan |B_\ell^N(\cdot)|, \nu_\ell^N\ran + \lan |B_\ell^N(\cdot)|^2, \nu_\ell^N\ran\big)\\
	&\le \f {C_2}N\sum_{N\t \le \ell \le N(\t + u)}\big(1 + \lan |B_\ell^N(\cdot)|^2, \nu_\ell^N\ran\big).
\end{align*}
Hence, applying Lemma \ref{lemma:bound_c_w} concludes the proof.
\enp

\section{Uniqueness }
\label{uniq_sec}
In this section, we show that Equation \eqref{evol_eq} admits a unique solution. For this we rely on a Picard-type argument for the ODE on $S$ of the form
\begin{align}
	\label{ode_eq}
	\frac{\d}{\d t} \th_t = A(\th_t; \mu_t),
\end{align}
with a generic $\mu_\cdot \in \DT$. Writing $D_T = D([0, T], S)$, this system gives rise to an operator 
$ H:\, \DTT \to \DTT$
as follows. First, if $\mu \in \DTT$ describes the distribution of a random path, then we let $\mu_0 \in \MM(S)$ denote the distribution of the initial point. Now, we define $H(\mu)$ to be the distribution of the solution $\{\theta_t\}_{t \le T}$  
to \eqref{ode_eq} with initial value distributed according to $\mu_0$.

%
%
The key observation is that $H$ has a unique fixed point if restricted to a smaller space. To introduce this space rigorously, we first put $C_T = C([0, T], S)$ and $M_T = \MM(C_T)$. Next, proceeding as in \cite[p.742]{sirignano.2018}, for $\mu, \mu' \in M_T$ let the \emph{coupling set} $P(\mu, \mu')$ denote the family of all probability measures on $C_T \times C_T$ coinciding with $\mu$ and $\mu'$ when projecting on the first and second marginal, respectively. Then,
 \begin{equation}\label{eq:wasserstein_def}
     d_{\ms W, T}(\mu, \mu') = \inf_{\nu \in P(\mu, \mu')}\Big(\int 1\wedge \sup_{s \le T}|u_s - v_s|^4_4 \nu(\d(u_\cdot, v_\cdot))\Big)^{1/4}
 \end{equation}
defines the \emph{$4$-Wasserstein} distance between $\mu$ and $\mu'$, where $|\cdot|_4$ denotes the $\ell^4$-distance in $S$. We write $N_T \su \MM(C([0, T], S))$ for the subspace of all $\mu \in M_T$ such that 
$\int\sup_{s \le T}|u_s|^4_4 \mu(\d u_\cdot) < \ff$, so that $N_T$ becomes a Banach space with respect to $d_{\ms W, T}$, see~\cite[p.743]{sirignano.2018}. 
\bel[Fixed point]
\label{fix_lem}
If $T$ is sufficiently small, then the restriction of $H$ to the space $N_T$ admits a unique fixed point.
\enl

%
%
\bep
Having set up the distance notion in~\eqref{eq:wasserstein_def}, we now show that $H$ is a contraction with respect to $d_{\ms W, T}$.
First, the evolution equation for $c_t$ does not change at all through our pruning, so that we can import the estimates from Lemma 4.3 in~\cite{sirignano.2018} to conclude that 
$$|c_t^{(1)} - c_t^{(2)}| \le C\int_0^t(|w_s^{(1)} - w_s^{(2)}| + d_{\ms W, s}(\mu^{(1)}, \mu^{(2)}))\d s$$
for a suitable $C > 0.$  Next, we decompose $w_t^{(1)} - w_t^{(2)}$ as
\begin{align*}
	w_t^{(1)} - w_t^{(2)} &=  \int_0^t\E\Big[X(\xi)(g(X, Y, \mu_s^{(1)}) - g(X, Y, \mu_s^{(2)}))  c^{(1)}_s\s'(w^{(1)}_s \cdot X) \Big] \d s\\
	&\phantom=+ \int_0^t\E\Big[X(\xi)g(X, Y, \mu_s^{(2)})  (c_s^{(1)}\s'(w^{(1)}_s \cdot X) - c_s^{(2)}\s'(w^{(2)}_s \cdot X))\Big] \d s.
\end{align*}
The only difference to the corresponding expression in Lemma 4.3 of~\cite{sirignano.2018} is that we now see $X(\xi)$ instead of $X$. However, in the ensuing estimates $X$ only appears through its length $|X|$. Since $|X(\xi)| \le |X|$, the arguments extend to the novel setting. Note that Lemma 4.3 in~\cite{sirignano.2018} requires that $\E \exp(q |c^i_0|) < \infty$ for some $ q > 0$. More precisely, this expression appears after an application of Gr\"{o}nwall's Lemma, see, e.g., Appendix 5 in~\cite{kurtz}.
\enp

%
%
In order to deduce Proposition \ref{uniq_prop} from Lemma \ref{fix_lem}, we need that solutions to \eqref{ode_eq} are indeed contained in $N_T$. 
\bel[Regularity of solutions]
\label{reg_ode_lem}
Let $\mu \in \DTT$ and let condition {\bf{(M)}} be fulfilled. Then, $H(\mu) \in N_T$. \enl
%
%
\bep
%
%
First, analogously to Lemma 4.1 in~\cite{sirignano.2018}, there exists a constant $C > 0$ such that $$\E[(c_t - c_s)^4] \le C (t-s)^4$$  and $$\E [ | w_t - w_s |^4 ] \le C ( \E [ |c_0|^4 ] + 1) (t-s)^4$$. These bounds imply that the processes $\{ c_t \}_{t \geq 0}$ and $\{ w_t \}_{t \geq 0}$ have continuous versions according to the Kolmogorov-Chentsov criterion, see Theorem 3.23 in~\cite{kallenberg.2002}. Moreover, they also imply that the solution curves have bounded fourth moments, so that indeed $H(\mu_\cdot) \in N_T$. 
\enp

%
%
Finally, we conclude the proof of Proposition \ref{uniq_prop}.
\bep[Proof of Proposition \ref{uniq_prop}]
We may choose $T$ to be small enough, so that Lemma \ref{fix_lem} applies. First, as in Section 4 of \cite{sirignano.2018}, general results on Markov processes from \cite{kolokoltsov} yield that solutions to \eqref{evol_eq} correspond uniquely to solutions of \eqref{ode_eq} by taking $\bar\mu_t$ to be the law of $\{\theta_t\}_{t \le T}$. In particular, the law of $\{\theta_t\}_{t \le T}$ is a fixed point of $H$ and therefore contained in $N_T$ by Lemma \ref{reg_ode_lem}. Hence, the uniqueness result from Lemma \ref{fix_lem} concludes the proof.
\enp

\section{Limit identification}
\label{identify_sec}
Last not least, we prove Proposition \ref{identify_prop}. That is, any limit point of the processes $\{\mu^N_t\}_t$ satisfies Equation \eqref{evol_eq}. Fix $f \in C_b^2(S)$. The central task is to quantify the error of $\lf{\mu_t^N} - \lf{\mu_0^N}$ in comparison to \eqref{evol_eq}.

%
%
\bel[Deviation from evolution equation]\label{dev_lem}
Let $t \le T$ and $f \in C_b^2(S)$. Then,
$$\De_t(\mu_\cdot^N) = \lf{ \mu_t^N} - \lf{ \mu_0^N} - \int_0^t \lan A(\cdot; \mu_s^N) \nabla f , \mu_s^N\ran \d s $$
converges to 0 in probability as $N \to \ff$.
\enl

%
%
First, we elucidate how to derive Proposition \ref{identify_prop} from Lemma \ref{dev_lem}.
\bep[Proof of Proposition \ref{identify_prop}]
Let $\Q$ be a weak accumulation point of $\{\mu_\cdot^N\}_{N \ge 1}$ and let $\mu_\cdot$ be a process distributed according to $\Q$. It suffices to prove that $\De_\cdot(\mu_\cdot) \equiv 0$ as a stochastic process, since then $\Q$ is concentrated on the unique solution of Equation~\eqref{evol_eq}. To that end, we verify that 
$	\E_{\Q}[\De_t(\mu_\cdot)G(\mu_\cdot)] = 0$,
for every $t > 0$ and bounded function $G:\mathcal{M}(D_T) \rightarrow [0, \infty)$ that is measurable with respect to $\{\mu_s\}_{s \le t}$. Since measurability is considered via the product $\s$-algebra, it suffices to fix arbitrary $s_1 < \cdots < s_p \le t$ and $g_1, \dots, g_p \in C_b(\R^{1 + d})$, and then show that
\begin{align}
	\label{qz_eq}
	\E_{\Q} \De'(\mu_\cdot) = 0,
\end{align}
where 
$\De'(\mu_\cdot) = \De_t(\mu_\cdot)\lan g_{s_1}, \mu_{s_1}\ran \cdots \lan g_{s_p}, \mu_{s_p}\ran.$ Now, since $\De'(\mu_\cdot)$ is bounded and continuous in $\mu_\cdot$, and $\Q$ is a weak accumulation point of a subsequence $\{\LL(\mu^{N_j})\}_{j \ge 1}$, we leverage Lemma \ref{dev_lem} to deduce that
\begin{align*}
	\E_{\Q} |\De'(\mu_\cdot)| \le \limsup_{j \to \ff} \E |\De'(\mu^{N_j})| \le   \max_i(|g_i|_\infty)\lim_{j \to \ff}\E |\De_t(\mu^{N_j})|  = 0,
\end{align*}
as asserted.
\enp

%
%
It remains to prove Lemma \ref{dev_lem}. 
\bep[Proof of Lemma \ref{dev_lem}]
By relying on a Taylor expansion as in the proof of Proposition \ref{reg_lem}, we see that 
$$\Big|\lf{\mu_t^N} - \lf{\mu_0^N}- \f1N\sum_{k \le Nt}\lan B_k(\cdot; \nu_k^N)\nabla f, \nu_k^N\ran\Big| \le \f {C_1}{N^2}\sum_{ k\le Nt} \lan |B_\ell^N(\cdot)|^2, \nu_\ell^N\ran$$ for some constant $ C_1 > 0.$
Now, as in Proposition \ref{reg_lem}, we deduce that the expression $$N^{-2} \, \E [\sum_{ k\le Nt} \lan |B_\ell^N(\cdot)|^2, \nu_\ell^N\ran ]$$ tends to 0 as $N \to\ff$. Thus, it suffices to show that 
\begin{align*}
	M(t) &= \f1N\sum_{k \le Nt}\lan B_k(\cdot; \nu_k^N)\nabla f, \nu_k^N\ran-  \int_0^t \lan A(\cdot; \mu_s^N) \nabla f , \mu_s^N\ran \d s \\
	&= \f1N\sum_{k \le Nt}\lan (B_k(\cdot; \nu_k^N) - A(\cdot; \nu_k^N))\nabla f, \nu_k^N\ran - \int_{\lfloor Nt \rfloor / N}^t \lan A(\cdot; \nu_{\lfloor Ns \rfloor}^N) \nabla f , \nu_{\lfloor Ns \rfloor}^N\ran \d s
\end{align*}
 tends to 0 in probability. We even show that it tends to 0 in $L^1$. Since 
 \begin{align*}
	 \E  | \lan A(\cdot; \nu_{\lfloor Ns \rfloor}^N))\nabla f , \nu_{\lfloor Ns \rfloor}^N\ran | \le   C_2\E \lan \E[|Y| + |g(X, \cdot)|],  \nu_{\lfloor Ns \rfloor}^N\ran 
	 \le C_3 + \frac{C_3}N \sum_{i \le N}  \E | c^i_{\lfloor Ns \rfloor}|
 \end{align*}
	 for some constants $C_2, C_3 > 0$, we obtain by \eqref{c_ell_eq} that the integral term of $M(t)$ tends to 0 in $L^1$. Moreover, we show that the sum appearing in the expression of $M(t)$ tends to $0$ in $L^2.$ By setting $M_k = N^{-1} \, \lan (B_k(\cdot; \nu_k^N) - A(\cdot; \nu_k^N))\nabla f, \nu_k^N\ran,$ we observe that since the training data $\{(X_k, Y_k)\}_{k \ge1}$ is iid, the sequence $\{M_k\}_{k \geq 1}$ defines a  martingale difference sequence. Therefore, the cross-terms in the expansion of the square disappear, i.e., $\sum_{k < k' \le Nt} \E[M_k M_{k'}] = 0$ and thus $\E \big(\sum_{k \le Nt} M_k\big)^2  = \sum_{k \le Nt} \E M_k^2.$ Noting that each summand $\E M_k^2$ is of order $1/N$ concludes the proof.
\enp

\newpage

\section*{Supplementary material}

\begin{figure}[h!]
    \centering
    \begin{tabular}{cc}
        \includegraphics[width = 0.44\textwidth]{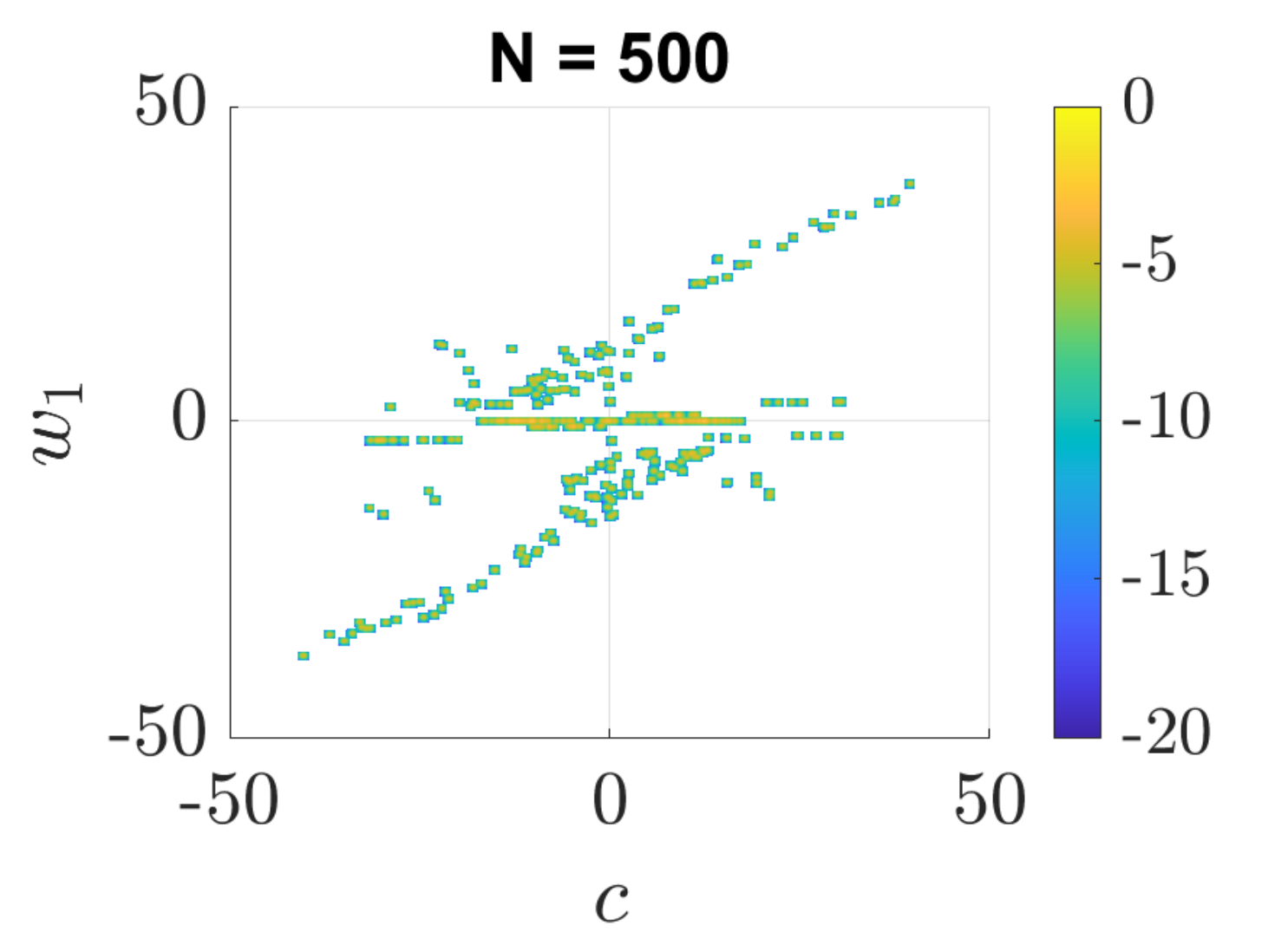} &  
        \includegraphics[width = 0.44\textwidth]{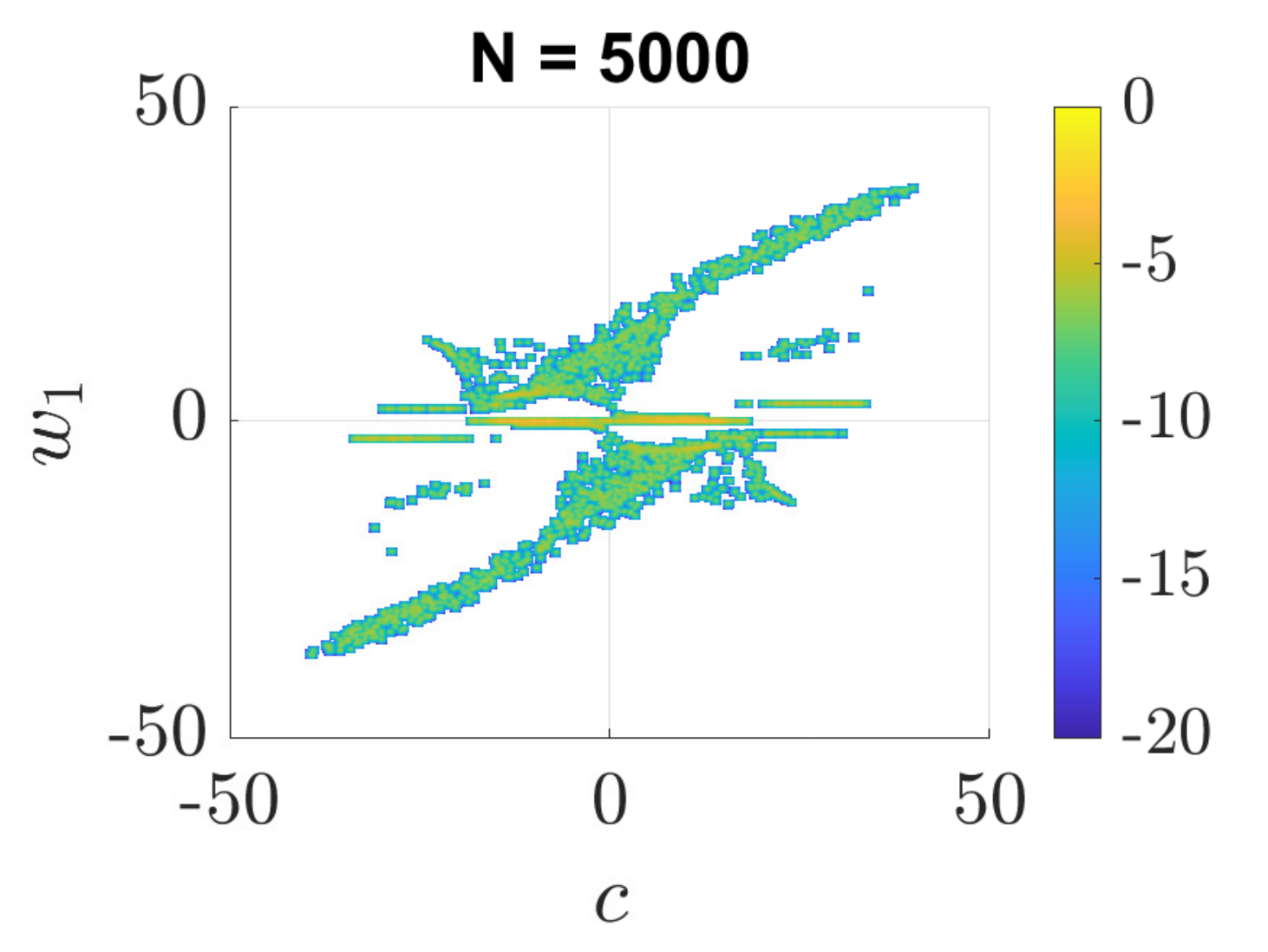} \\
        \includegraphics[width = 0.44\textwidth]{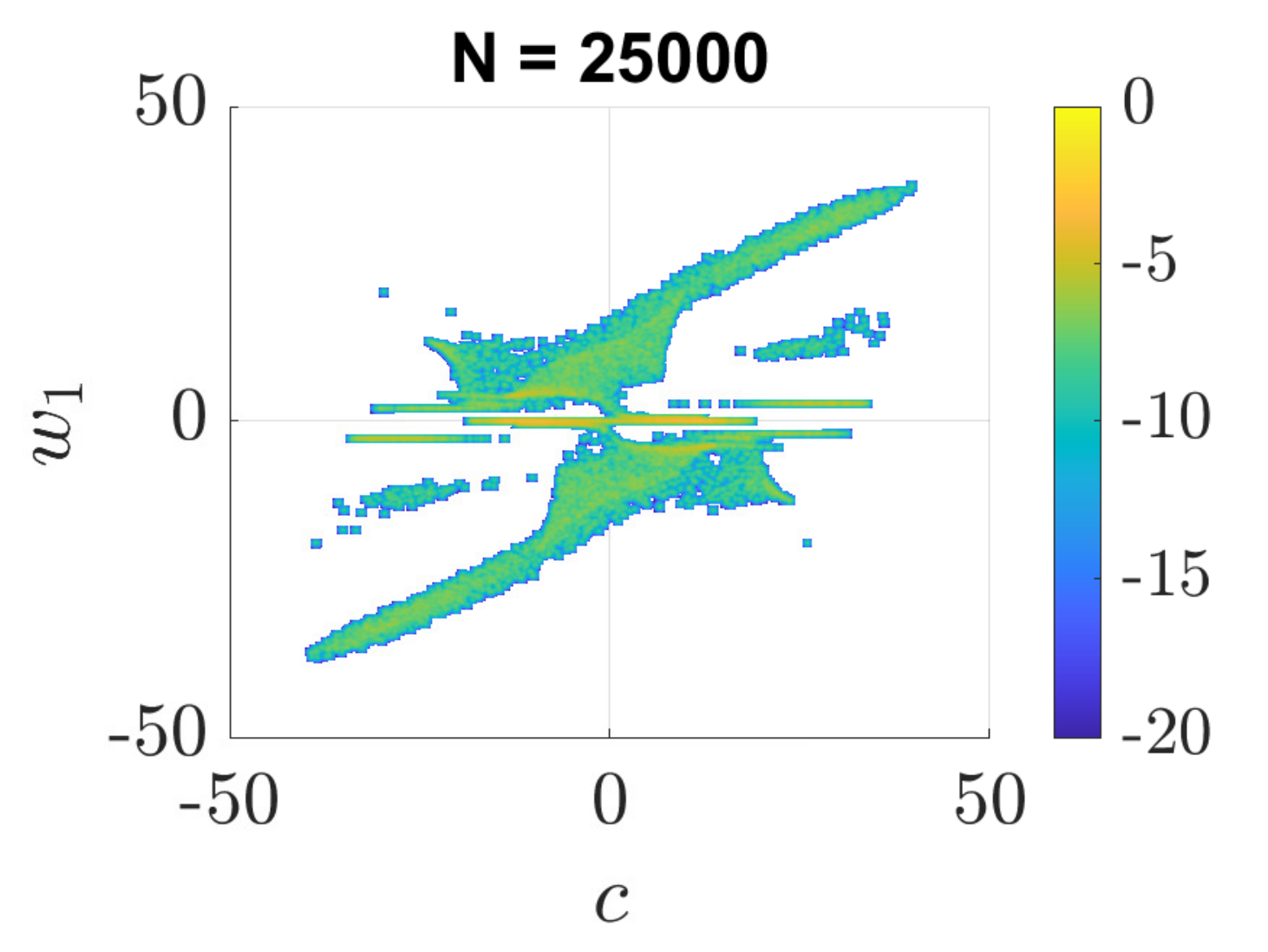} &  
        \includegraphics[width = 0.44\textwidth]{images/CW1Bivariate3-eps-converted-to.pdf} 
    \end{tabular}
    
    \caption{Probability density function of the parameter vector $(c_{KN}, (w_{KN})_1)$ after $KN$ iterations of the SGD for $N \in \lbrace 500, 5000, 25000, 50000 \rbrace,$ obtained via kernel density estimation. The values of the density are represented on the log-scale, where domains which do not belong to the support of the estimated distribution are represented in white.}
    \label{fig:cw1}
\end{figure}

\begin{figure}[h!]
    \centering
    \begin{tabular}{cc}
        \includegraphics[width = 0.44\textwidth]{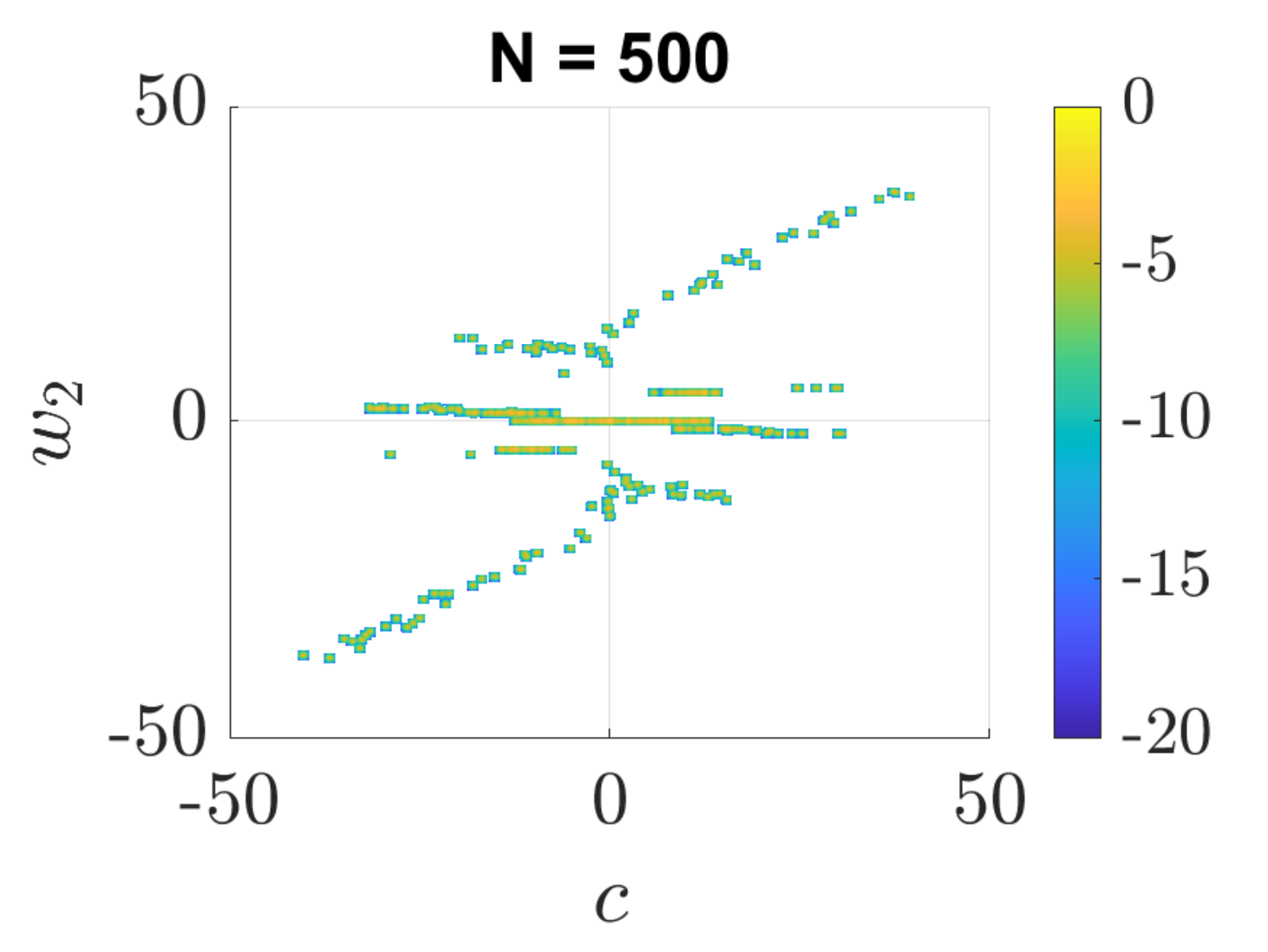} &  
        \includegraphics[width = 0.44\textwidth]{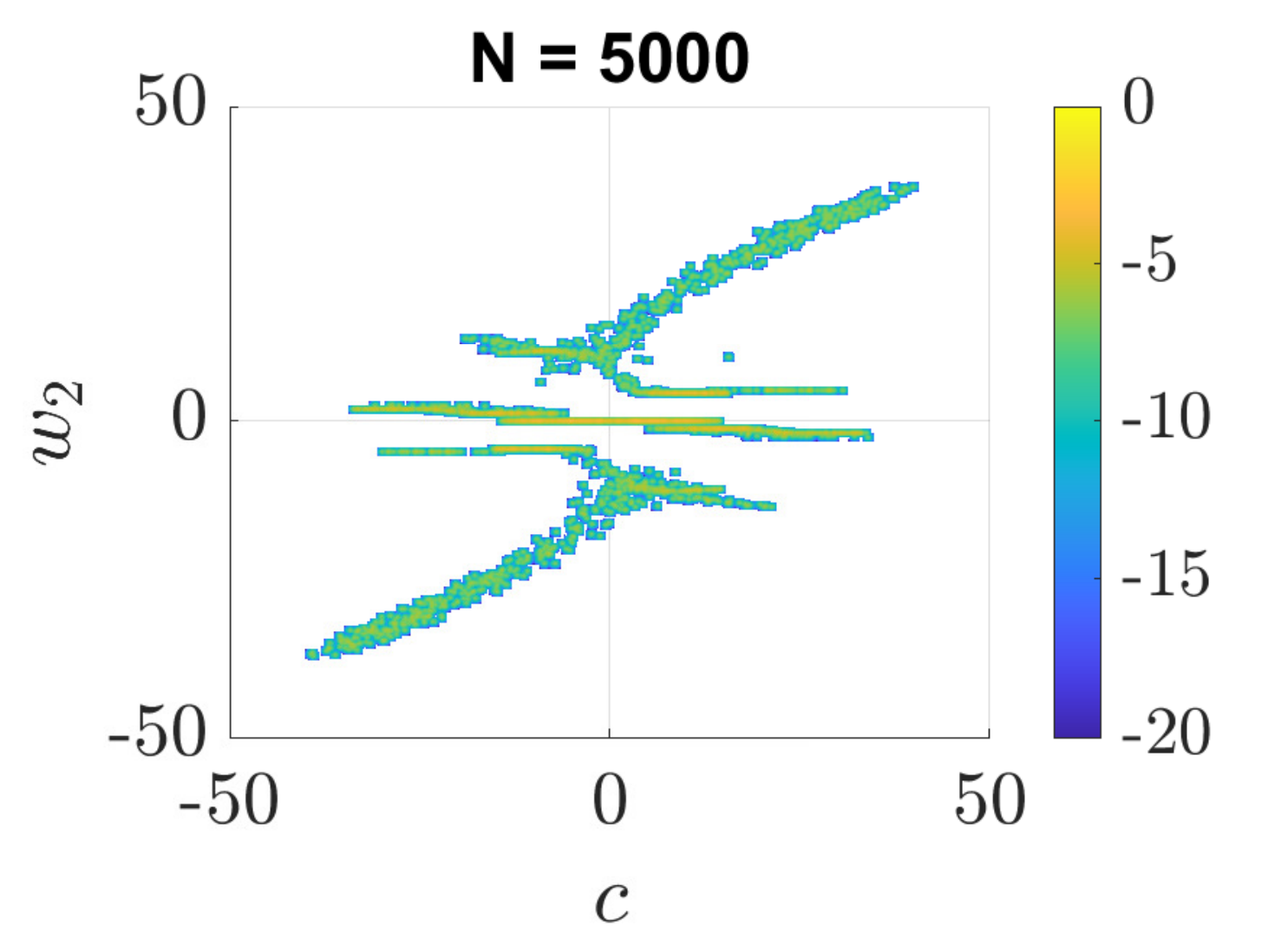} \\
        \includegraphics[width = 0.44\textwidth]{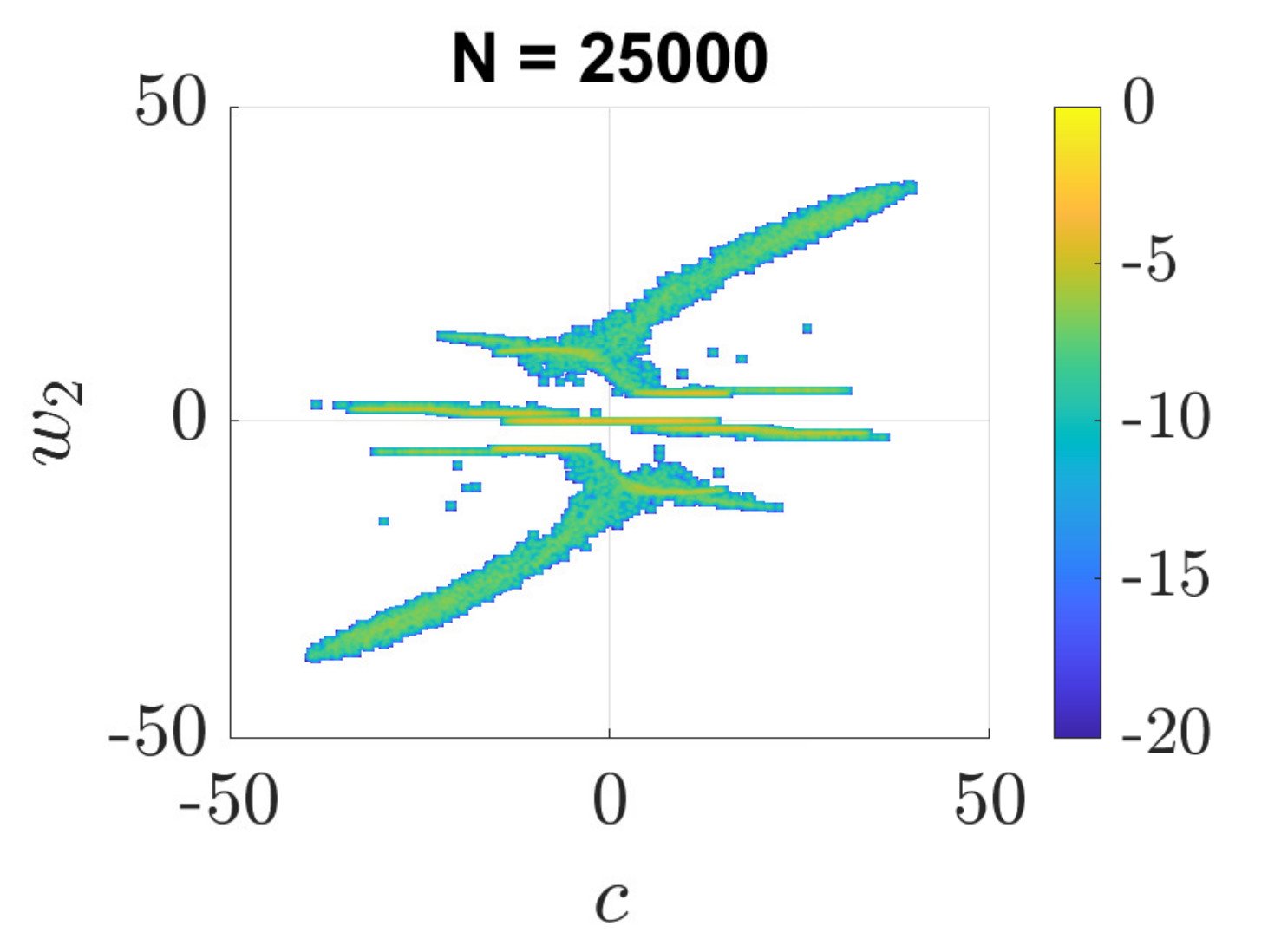} &  
        \includegraphics[width = 0.44\textwidth]{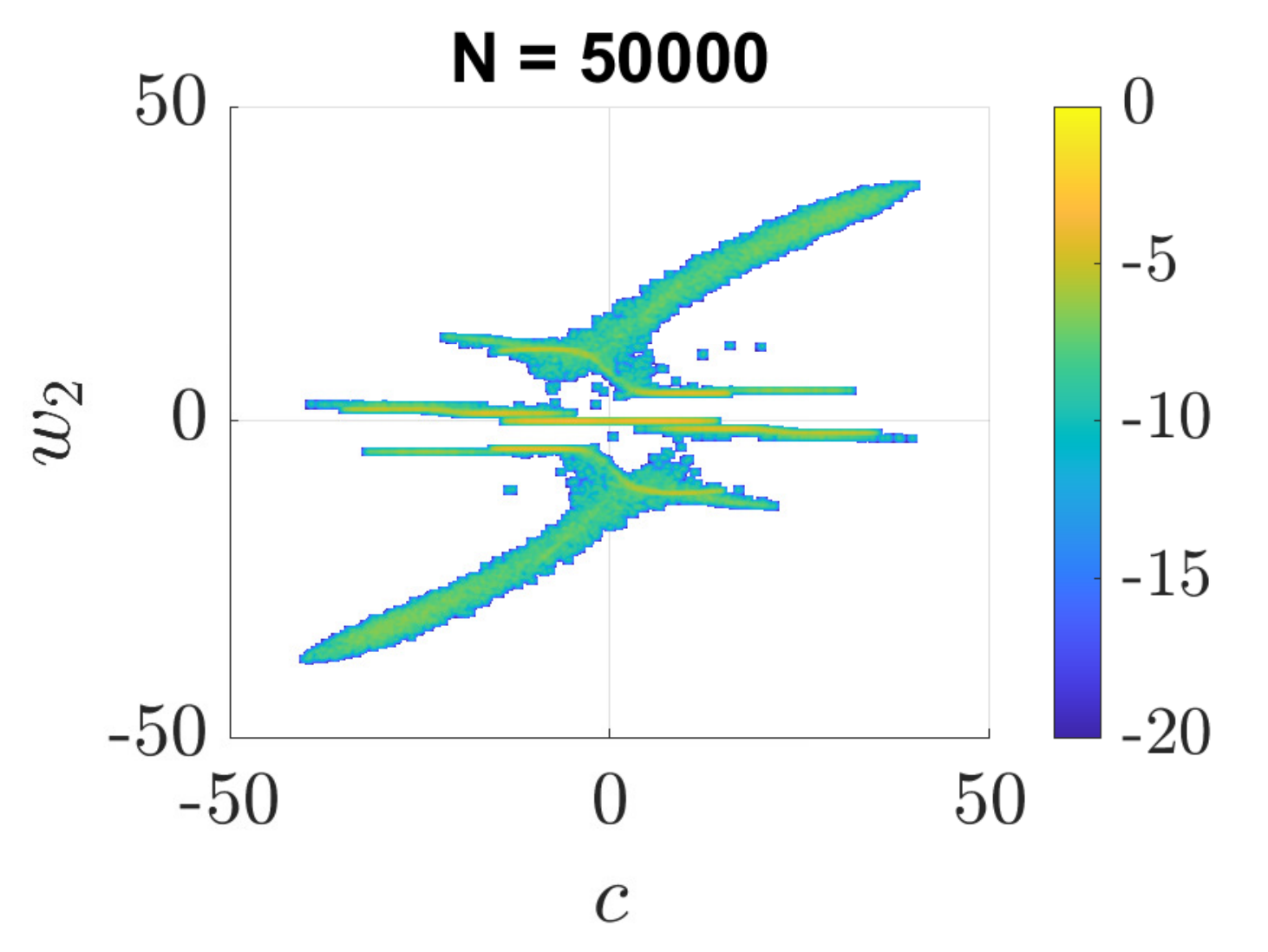} 
    \end{tabular}
    
    \caption{Probability density function of the parameter vector $(c_{KN}, (w_{KN})_2)$ after $KN$ iterations of the SGD for $N \in \lbrace 500, 5000, 25000, 50000 \rbrace,$ obtained via kernel density estimation. The values of the density are represented on the log-scale, where domains which do not belong to the support of the estimated distribution are represented in white.}
    \label{fig:cw2}
\end{figure}

\begin{figure}[h!]
    \centering
    \begin{tabular}{cc}
        \includegraphics[width = 0.44\textwidth]{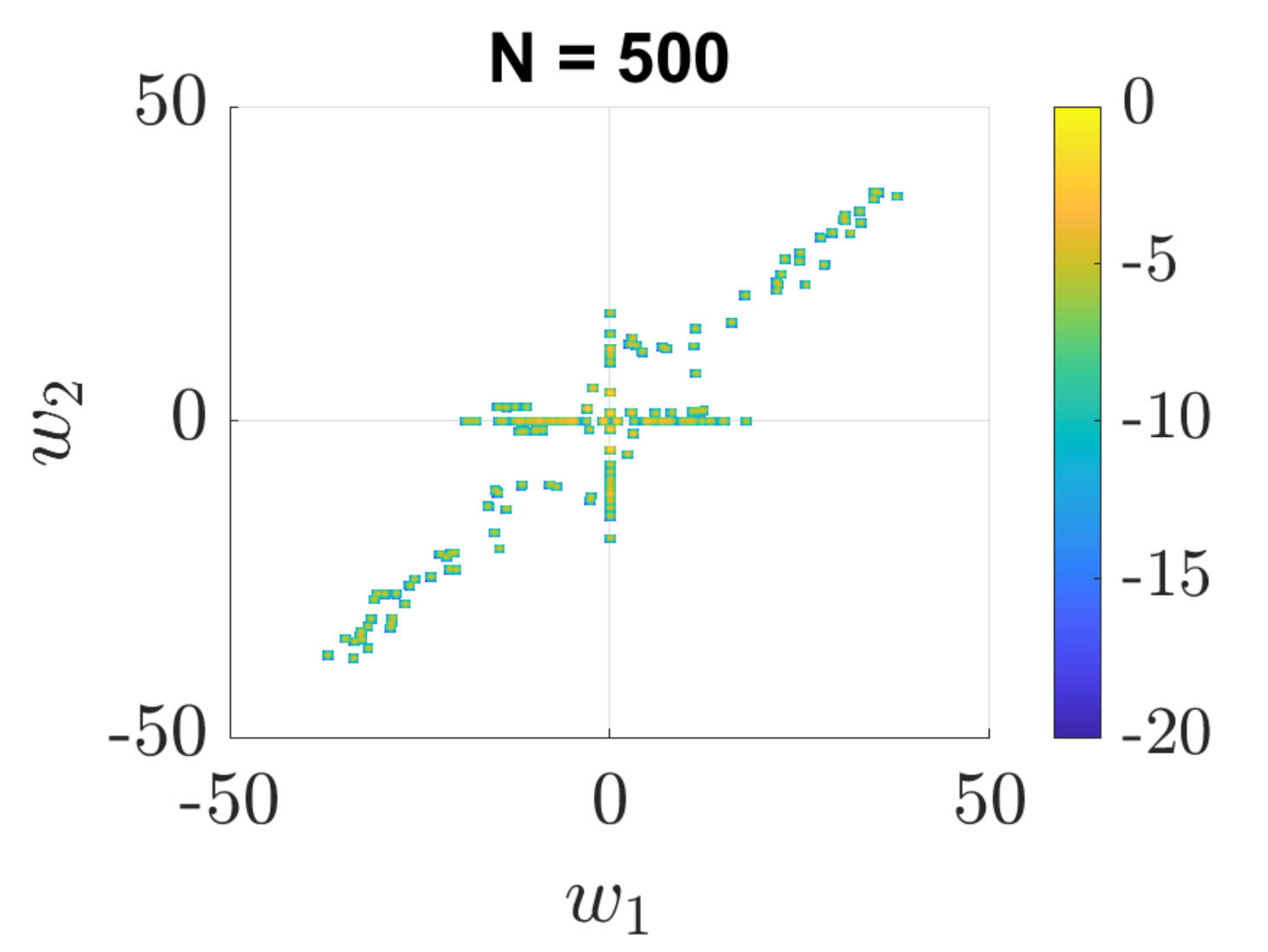} &  
        \includegraphics[width = 0.44\textwidth]{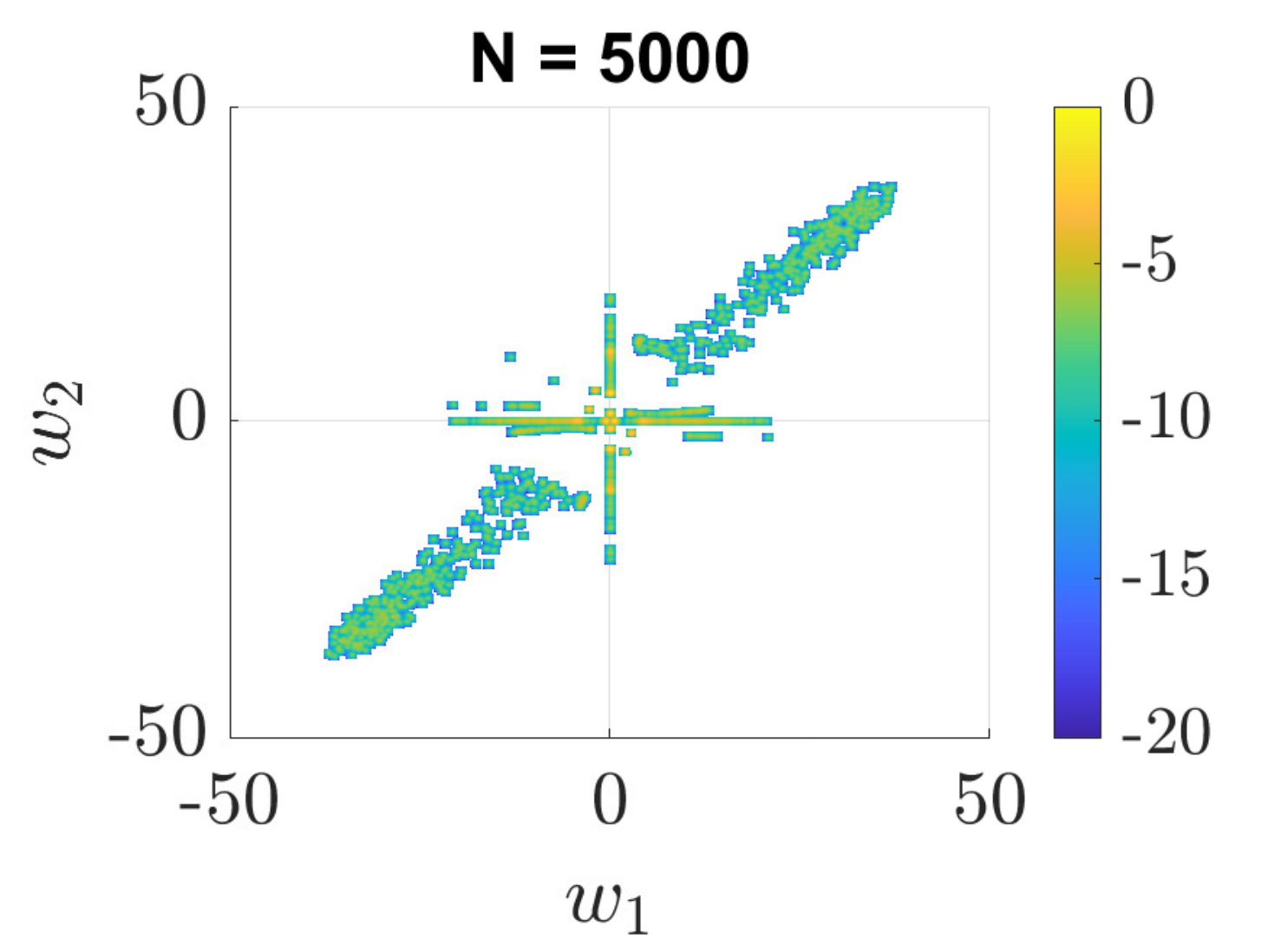} \\
        \includegraphics[width = 0.44\textwidth]{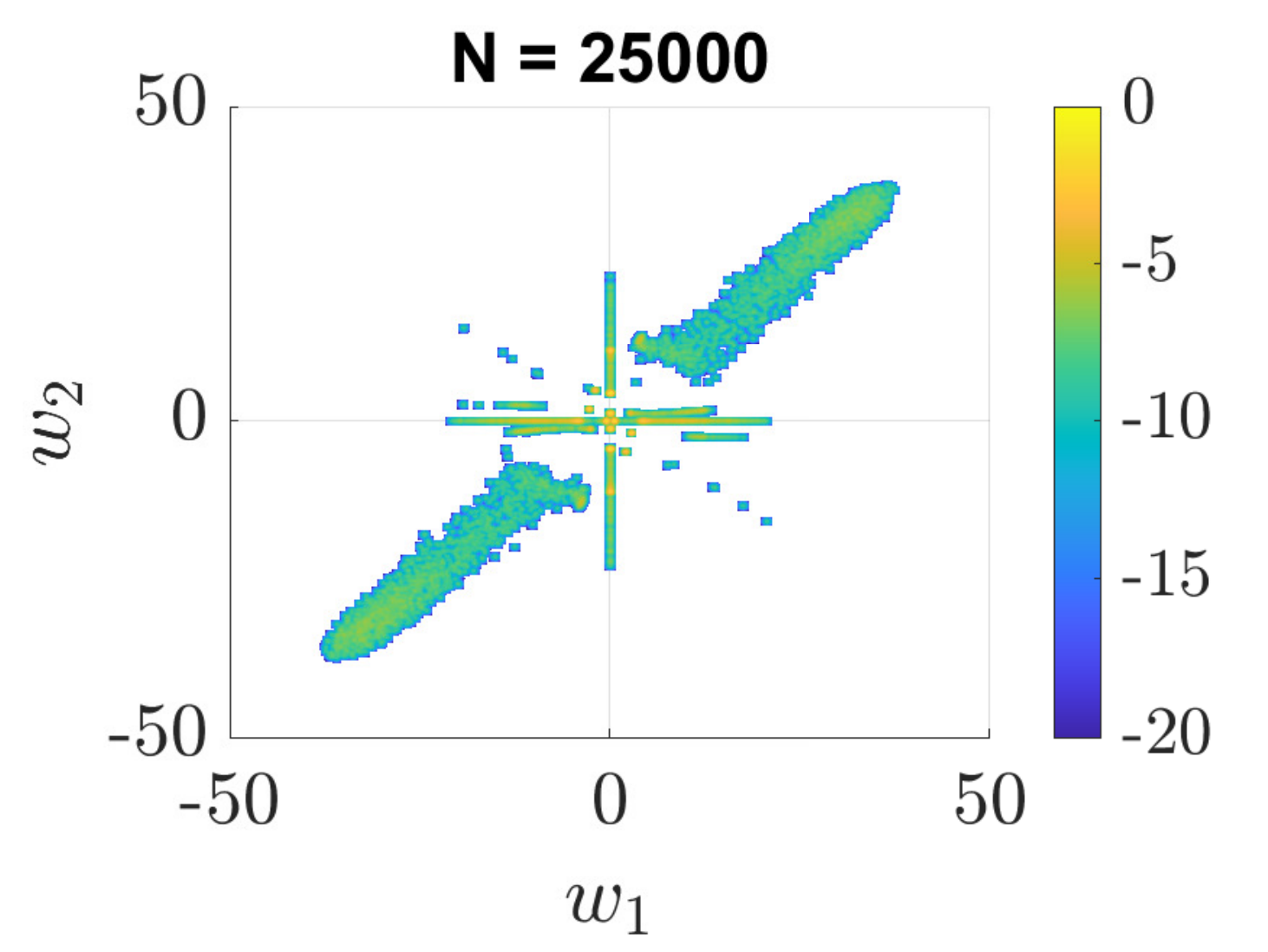} &  
        \includegraphics[width = 0.44\textwidth]{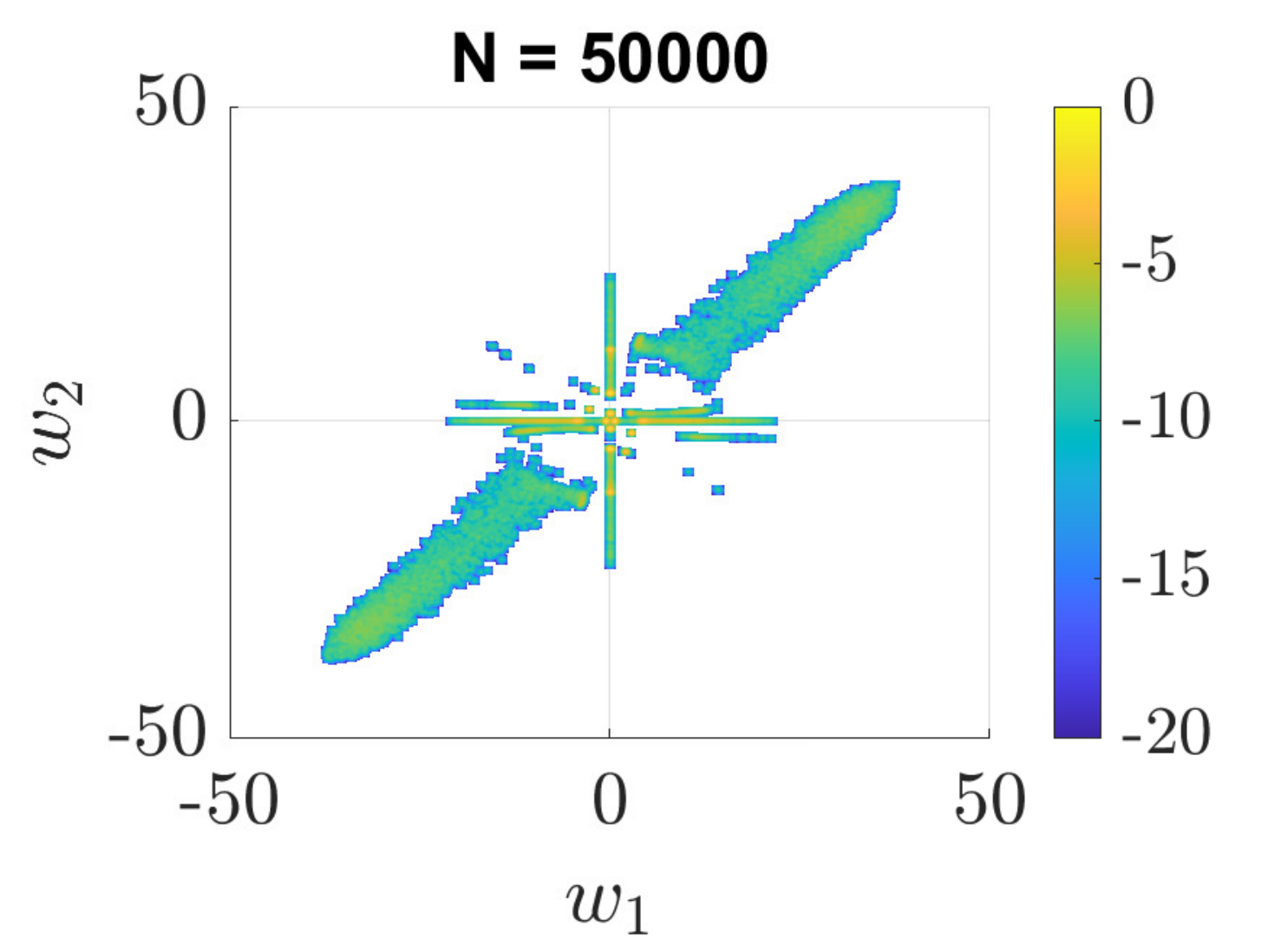} 
    \end{tabular}
    
    \caption{Probability density function of the parameter vector $((w_{KN})_1, (w_{KN})_2)$ after $KN$ iterations of the SGD for $N \in \lbrace 500, 5000, 25000, 50000 \rbrace,$ obtained via kernel density estimation. The values of the density are represented on the log-scale, where domains which do not belong to the support of the estimated distribution are represented in white.}
    \label{fig:w1w2}
\end{figure}

As supplementary material, we provide further results related to the simulation study presented in Section~\ref{sec:simulation}. We show the empirical bivariate distributions of the samples $(c^1_{KN}, (w^1_{KN})_1),$ $\ldots, (c^N_{KN}, (w^N_{KN})_1)$, $(c^1_{KN}, (w^1_{KN})_2), \ldots, (c^N_{KN}, (w^N_{KN})_2)$, and $((w^1_{KN})_1), (w^1_{KN})_2), \ldots,$ \\$((w^1_{KN})_1), (w^N_{KN})_2)$ for $N = 500, N = 5000, N = 25000, N = 50000$ in Figures~\ref{fig:cw1},~\ref{fig:cw2},~\ref{fig:w1w2}, respectively. The distributions are shown in terms of probability density functions estimated by kernel density estimation. For this purpose, a bivariate Gaussian kernel with a bandwidth of 0.2 is used. The values of the estimated probability density functions are represented by a heat map on the log-scale. Domains which do not belong to the support of the estimated probability density function are represented in white. Figures~\ref{fig:cw1},~\ref{fig:cw2},~\ref{fig:w1w2} nicely show how the considered empirical bivariate distributions approach the limit distribution with increasing values of $N$.

\end{document}